\documentclass[%linenumbers, 
twocolumn, 
astrosymbooll]{aastex631}

\hypersetup{linkcolor=blue,citecolor=blue,filecolor=cyan,urlcolor=blue}

\usepackage{booktabs}

\newcommand\eg{{e.g.}}

\usepackage{longtable}
\usepackage{graphicx}%Include figure files
\usepackage[normalem]{ulem}
\usepackage{ulem}

\usepackage{multirow}
\usepackage{url}

\defcitealias{COSEP}{COSEP}
\defcitealias{lsst}{I19}
\defcitealias{jones2020}{PSTN-051}

\newcommand{\maf}{\texttt{MAF}\,}

\newcommand{\ie}{\emph{i.e.}}
\newcommand{\opsim}{\texttt{OpSim}\,}

\shorttitle{Rubin Observatory Legacy Survey of Space and Time Observing Strategy}
\shortauthors{Bianco et al.}
\begin{document}

\title{Optimization of the Observing Cadence for the Rubin Observatory Legacy Survey of Space and Time: a pioneering process of community-focused experimental design}

\author[0000-0003-1953-8727]{Federica B. Bianco}
\affiliation{University of Delaware
Department of Physics and Astronomy
217 Sharp Lab
Newark, DE 19716 USA}
\affiliation{University of Delaware
Joseph R. Biden, Jr. School of Public Policy and Administration, 
184 Academy St, Newark, DE 19716 USA}
\affiliation{University of Delaware
Data Science Institute}
\affiliation{Center for Urban Science and Progress, New York University, 
370 Jay St, Brooklyn, NY 11201, USA}

\author[0000-0001-5250-2633]{\v{Z}eljko Ivezi\'{c}}
\affiliation{Department of Astronomy and the DiRAC Institute, University of Washington, 3910 15th Avenue NE, Seattle, WA 98195, USA}

\author[0000-0001-5916-0031]{R. Lynne Jones}
\affiliation{Aerotek and Rubin Observatory, Tucson}

\author[0000-0002-9154-3136]{Melissa L. Graham}
\affiliation{Department of Astronomy and the DiRAC Institute, University of Washington, 3910 15th Avenue NE, Seattle, WA 98195, USA}

\author{Phil Marshall}
\affiliation{SLAC National Accelerator Laboratory, Menlo Park, CA 94025, USA}

\author[0000-0002-6839-4881]{Abhijit Saha}
\affiliation{NSF’s National Optical-Infrared Astronomy Research Laboratory, 950 N. Cherry Ave., Tucson, AZ 85719, USA}

\author[0000-0002-0106-7755]{Michael A. Strauss}
\affiliation{Department of Astrophysical Sciences, Princeton University, Princeton, NJ 08544 USA}

\author[0000-0003-2874-6464]{Peter Yoachim}
\affiliation{Department of Astronomy, University of Washington, 3910 15th Avenue NE, Seattle, WA 98195, USA}

\author[0000-0002-0138-1365]{Tiago Ribeiro}
\affiliation{LSST Project Office, 950 N.\ Cherry Ave., Tucson, AZ  85719, USA}

\author[0000-0003-0930-5815]{Timo Anguita}
\affiliation{Departamento de Ciencias Fisicas, Universidad Andres Bello Fernandez Concha 700, Las Condes, Santiago, Chile}
\affiliation{Millennium Institute of Astrophysics, Nuncio Monse{\~{n}}or S{\'{o}}tero Sanz 100, Of 104, Providencia, Santiago, Chile} 

\author[0000-0002-8686-8737]{Franz E. Bauer}
\affiliation{Instituto de Astrof{\'{\i}}sica and Centro de Astroingenier{\'{\i}}a, Facultad de F{\'{i}}sica, Pontificia Universidad Cat{\'{o}}lica de Chile, Casilla 306, Santiago 22, Chile} 
\affiliation{Millennium Institute of Astrophysics, Nuncio Monse{\~{n}}or S{\'{o}}tero Sanz 100, Of 104, Providencia, Santiago, Chile} 
\affiliation{Space Science Institute, 4750 Walnut Street, Suite 205, Boulder, Colorado 80301, USA} 

\author[0000-0001-8018-5348]{Eric C. Bellm}
\affiliation{DiRAC Institute, Department of Astronomy, University of Washington, 3910 15th Avenue NE, Seattle, WA 98195, USA}

\author[0000-0002-8622-4237]{Robert D. Blum}
\affiliation{Vera C. Rubin Observatory/NSF's NOIRLab, 950 North Cherry Avenue, Tucson, AZ, 85719, USA}

\author[0000-0002-0167-2453]{William N. Brandt}
\affiliation{Department of Astronomy and Astrophysics, 525 Davey Lab, The Pennsylvania State University, University Park, PA 16802, USA}
\affiliation{Institute for Gravitation and the Cosmos, The Pennsylvania State University, University Park, PA 16802, USA
}
\affiliation{Department of Physics, 104 Davey Laboratory, The Pennsylvania State University, University Park, PA 16802, USA}

\author[0000-0002-9796-1363]{Sarah Brough}
\affiliation{ARC Centre of Excellence for All Sky Astrophysics in 3 Dimensions (ASTRO 3D), Australia}
\affiliation{School of Physics, University of New South Wales, NSW 2052, Australia}

\author[0000-0001-6003-8877]{Márcio Catelan}
\affiliation{Millennium Institute of Astrophysics, Nuncio Monse{\~{n}}or S{\'{o}}tero Sanz 100, Of 104, Providencia, Santiago, Chile} 
\affiliation{Instituto de Astrofísica, Pontificia Universidad Católica de Chile, Av. Vicuña Mackenna 4860, 7820436 Macul, Santiago, Chile}
\affiliation{Centro de Astro-Ingeniería, Pontificia Universidad Católica de Chile, Av. Vicuña Mackenna 4860, 7820436 Macul, Santiago, Chile}

\author[0000-0002-2577-8885]{William I. Clarkson}
\affiliation{Department of Natural Sciences, University of Michigan - Dearborn, 4901 Evergreen Road, Dearborn, MI 48128, USA}

\author[0000-0001-5576-8189]{Andrew J. Connolly}
\affiliation{Department of Astronomy and the DiRAC Institute, University of Washington, 3910 15th Avenue NE, Seattle, WA 98195, USA}

\author[0000-0003-1530-8713]{Eric Gawiser}
\affiliation{Department of Physics and Astronomy, Rutgers, The State University of New Jersey, 136 Frelinghuysen Rd, Piscataway, NJ 08854, USA}

\author[0000-0002-8916-1972]{John Gizis}
\affiliation{University of Delaware
Department of Physics and Astronomy
217 Sharp Lab
Newark, DE 19716 USA}

\author[0000-0002-0965-7864]{Renee Hlozek}
\affiliation{Canadian Institute for Theoretical Astrophysics, University of Toronto, 60 St. George St., Toronto, ON M5S 3H4,
Canada}
\affiliation{Dunlap Institute of Astronomy \& Astrophysics, 50 St. George St., Toronto, ON M5S 3H4, Canada}

\author{Sugata Kaviraj}
\affiliation{Centre for Astrophysics Research, Department of Physics, Astronomy and Mathematics, University of Hertfordshire, Hatfield, AL10 9AB, UK}

\author[0000-0002-4314-8713]{Charles T. Liu}
\affiliation{Department of Physics \& Astronomy, City University of New York, College of Staten Island, Staten Island, NY 10314, USA}
\affiliation{Department of Astrophysics {\&} Hayden Planetarium, American Museum of Natural History,\\ \ New York, NY 10024-5192, USA}
\affiliation{Physics Program, The Graduate Center, CUNY, New York, NY 10016, USA}

\author[0000-0003-2221-8281]{Michelle Lochner}
\affiliation{Department of Physics and Astronomy, University of the Western Cape, Bellville, Cape Town, 7535, South Africa }
\affiliation{South African Radio Astronomy Observatory (SARAO), The Park, Park Road, Pinelands, Cape Town 7405, South Africa}

\author[0000-0003-2242-0244]{Ashish~A.~Mahabal}
\affiliation{Division of Physics, Mathematics and Astronomy, California Institute of Technology, Pasadena, CA 91125, USA}
\affiliation{Center for Data Driven Discovery, California Institute of Technology, Pasadena, CA 91125, USA}

\author[0000-0003-2271-1527]{Rachel Mandelbaum}
\affiliation{McWilliams Center for Cosmology, Department of Physics, Carnegie Mellon University, 5000 Forbes Ave, Pittsburgh, PA 15213}

\author[0000-0003-0948-6716]{Peregrine McGehee}\affiliation{College of the Canyons, 26455 Rockwell Canyon Rd, Santa Clarita, CA 91355}

\author[0000-0002-7357-0317]{Eric H. Neilsen, Jr.}
\affiliation{Fermi National Accelerator Laboratory, P.O. Box 500, Batavia, IL 60510, USA}

\author[0000-0002-7134-8296]{Knut A. G. Olsen}
\affiliation{NSF’s National Optical-Infrared Astronomy Research Laboratory, 950 N. Cherry Ave., Tucson, AZ 85719, USA}

\author[0000-0002-2519-584X]{Hiranya Peiris}
\affiliation{The Oskar Klein Centre for Cosmoparticle Physics, Department of Physics, Stockholm University, AlbaNova, Stockholm, SE-106 91,
Sweden}
\affiliation{Department of Physics \& Astronomy, University College London, Gower Street, London WC1E 6BT, UK}

\author[0000-0002-4485-8549]{Jason Rhodes}
\affiliation{Jet Propulsion Laboratory, California Institute of Technology, 4800 Oak Grove Drive, Pasadena, CA, 91109, USA}

\author[0000-0002-1061-1804]{Gordon T. Richards}
\affiliation{Department of Physics, Drexel University, 32 S.\ 32nd Street, Philadelphia, PA 19104, USA}

\author[0000-0003-2557-7132]{Stephen Ridgway}
\affiliation{NSF’s National Optical-Infrared Astronomy Research Laboratory, 950 N. Cherry Ave., Tucson, AZ 85719, USA}

\author[0000-0003-4365-1455]{Megan E. Schwamb}
\affiliation{Astrophysics Research Centre, School of Mathematics and Physics, Queen's University Belfast, Belfast BT7 1NN, UK}

\author{Dan Scolnic}
\affiliation{Department of Physics, Duke University, Durham, NC 27708, USA}

\author[0000-0003-4327-1460]{Ohad Shemmer}
\affiliation{Department of Physics, University of North Texas, Denton, TX 76203, USA}

\author[0000-0002-0558-0521]{Colin T. Slater}
\affiliation{DiRAC Institute and the Department of Astronomy, University of Washington,
Seattle, WA, U.S.A}

\author[0000-0002-8713-3695
]{An\v{z}e Slosar}
\affiliation{Physics Department, Brookhaven National Laboratory, Upton NY 11973}

\author[0000-0002-8229-1731]{Stephen J. Smartt}
\affiliation{Astrophysics Research Centre, School of Mathematics and Physics, Queen's University Belfast, Belfast BT7 1NN, UK}

\author[0000-0002-1468-9668]{Jay Strader}
\affiliation{Center for Data Intensive and Time Domain Astronomy, Department of Physics and Astronomy, Michigan State University, East Lansing MI, USA}

\author[0000-0001-6279-0552]{Rachel Street}
\affiliation{Las Cumbres Observatory, 6740 Cortona Drive, Suite 102,93117 Goleta, CA, USA}

\author[0000-0003-4580-3790]{David E. Trilling}
\affiliation{Department of Astronomy and Planetary Science
PO Box 6010
Northern Arizona University
Flagstaf, AZ 86011}

\author[0000-0002-0730-0781]{Aprajita Verma}
\affiliation{Sub-department of Astrophysics, Denys Wilkinson Building, University of Oxford, Keble Road, Oxford, OX2 6QX, U.K.}

\author[0000-0003-4341-6172]{A.~K.~Vivas}
\affiliation{Cerro Tololo Inter-American Observatory, NSF's National Optical-Infrared Astronomy Research Laboratory, Casilla 603, La Serena, Chile}

\author[0000-0003-2229-011X]{Risa H. Wechsler}
\affiliation{Department of Physics, Stanford University, 382 Via Pueblo Mall, Stanford, CA 94305, USA}
\affiliation{Kavli Institute for Particle Astrophysics and Cosmology, Stanford University, Stanford, CA 94305, USA}
\affiliation{Department of Particle Physics and Astrophysics, SLAC National Accelerator Laboratory, Stanford, CA 94305, USA}

\author[0000-0003-2892-9906]{Beth Willman}
\affiliation{NSF's NOIRLab, 950 North Cherry Avenue, Tucson, AZ 85721, USA}

\begin{abstract}
%HOW TO READ: \new{Blue is new and should be read over (but also hyperlinks are in blue)}, \question{Red is possibly wrong, or a question that needs to be answered, or a place holder. Should definitely be read}. \question{NOTE: the abstract needs to be $<250$ words}
Vera C. Rubin Observatory is a ground-based astronomical facility under construction, a joint project of the National Science Foundation and the U.S. Department of Energy, designed to conduct a multi-purpose 10-year optical survey of the southern hemisphere sky: the Legacy Survey of Space and Time. 
Significant flexibility in survey strategy remains within the constraints imposed by the core science goals of probing dark energy and dark matter, cataloging the Solar System, exploring the transient optical sky, and mapping the Milky Way. The survey's massive data throughput will be transformational for many other astrophysics domains and Rubin's data access policy sets the stage for a huge potential users' community. 
To ensure that the survey science potential is maximized while serving as broad a community as possible, Rubin Observatory has involved
the scientific community at large in the process of setting and refining the details of the observing strategy. The motivation, history, and decision-making process of this strategy optimization are detailed in this paper, giving context to 
the science-driven proposals and recommendations for the survey strategy included in this Focus Issue.
\end{abstract}



\section{Introduction} \label{sec:intro}

 Vera C. Rubin Observatory is designing an astronomical survey, the Legacy Survey of Space and Time (LSST) that will be revolutionary in many ways. The amount of imaging data, and the combination of flux sensitivity, area, and the temporal sampling rate will be dramatically increased compared to precursor surveys at any waveband. Rubin Observatory will observe the southern hemisphere sky from  El Peñón peak of Cerro Pachón, Chile, for 10 years, using the $8.36$~m aperture Simonyi Survey Telescope and the LSST Camera with its unique 9.6 deg$^2$ field of view to collect over two million sky images. It will do so with multiple filters ($ugrizy$) and with exquisite sub-arcsecond image quality. 
 
 Rubin Observatory's enormous dataset, which will reach a size of $\sim300$~PB when the 10 year survey ends, will be released with a unique 
 %open (Blum: I would not describe our policy as open)
 data policy. %Apart from proprietary data products,
Every night, millions of LSST alert packets that identify astrophysical transient, variable, and moving objects will be released worldwide in real time with no restrictions. 
The LSST images and catalogs will be available in their entirety 
%with no proprietary period 
to all Rubin data rights holders: any US and Chilean scientist and members of International Groups that have agreements with the National Science Foundation (NSF) and/or the Department of Energy (DOE) or their managing partners the Association of Universities for Research in Astronomy (AURA) and SLAC National Accelerator Lab, respectively.\footnote{See the Rubin Data Policy, \url{ls.st/rdo-013}, for more information about data rights holders}. 
These include catalogs available within 24 hours of observations, and annual releases of reprocessed images, deep coadded stacks, and associated catalogs\footnote{See the LSST Data Products Definitions Document, \url{ls.st/lse-163}, for more information about the LSST data products.}. These data will become available worldwide after two years from the original data release.
This enormous data set, combined with an open data access policy, sets the stage for a broad and diverse user community and a commensurately huge opportunity for maximum scientific impact from the LSST.

 LSST is designed to enable the pursuit of four main science themes: {probing dark energy and dark matter, exploring the transient optical sky,  mapping the Milky Way Galaxy, and building a catalog of Solar System objects over an order of magnitude larger than presently available}. These science goals were chosen to drive the survey design such that a transformational survey addressing all these science themes will also be able to impact
 %have the potential to being 
 %impactful in 
 many other fields of astrophysics, including, for example, galaxy morphology and evolution, AGN, and quasar studies, etc.  LSST science drivers and technical design details are summarized in the LSST overview paper \citep{lsst}.%\footnote{also available at \url{https://ls.st/lop}}.
 %One further aspect that makes this a unique project is that 

Rubin Observatory is constructing a flexible scheduling system that can respond to the unexpected and be re-optimized as the survey progresses. LSST observations will be scheduled automatically, with the scheduling algorithm designed to maximize scientific return under a set of observing constraints.  LSST is in fact an umbrella term to refers to a set of surveys that Rubin will perform in its first 10 years: a ``main survey" (hereafter referred to as Wide-Fast-Deep, WFD), a series of pointings that will be observed with an intensified cadence leading to deeper coadded image stacks, the Deep Drilling Fields (DDFs), and additional ``minisurveys'' that cover specific sky regions such as the ecliptic plane, Galactic plane, and the Large and Small Magellanic
Clouds, or that vary survey parameters such as the depth of a single visit. 

Any implementation of LSST’s 10-year observing strategy must meet the basic requirements described
in the LSST Science Requirements Document (SRD\footnote{The SRD document, available at \url{https://ls.st/srd}, indicates 3 levels of requirements: the design requirements, stretch goal, and minimum requirements. Here we report the design requirements.}) to ensure that the core science goals will be achieved: %(constraining dark energy and dark matter; taking an inventory of the Solar System; exploring the transient optical sky, and mapping the Milky Way; for a more detailed discussion of these science goals, please see \citealt{lsst}). 
a footprint for the WFD of at least 18,000 deg$^2$ which must be uniformly covered to a 
median of 825 nominal 30-second visits per 9.6 deg$^2$ field, summed over all six filters. 
Additional constraints on the temporal distribution of these visits are derived from requirements on parallax 
and proper motion accuracy, as well as a requirement for two visits per night to enable 
discovery of main-belt asteroids using standard algorithms. Yet, the SRD 
intentionally places minimal quantitative constraints 
on the observing strategy, recognizing that science evolves and the LSST plan can be refined until first light and even beyond. To that end, the Rubin Observatory operations 
team plans to continuously monitor survey progress and, if needed, to modify the strategy to achieve the desired scientific goals (see \autoref{sec:cn}).  
The SRD leaves significant flexibility in the detailed cadence of observations within the main survey footprint, including the distribution of 
visits within a year, the distribution of images between filters, and the definition of a 
``visit" itself (\eg, a single exposure or multiple exposures per visit). 
Furthermore, these constraints apply to the WFD only. Depending on the performance efficiency of the WFD, between 10\% and 20\% of the sky time will remain available for DDFs and minisurveys, whose design is not strongly prescribed by the SRD.

\begin{figure*}
\centering
\includegraphics[width=0.45\textwidth]{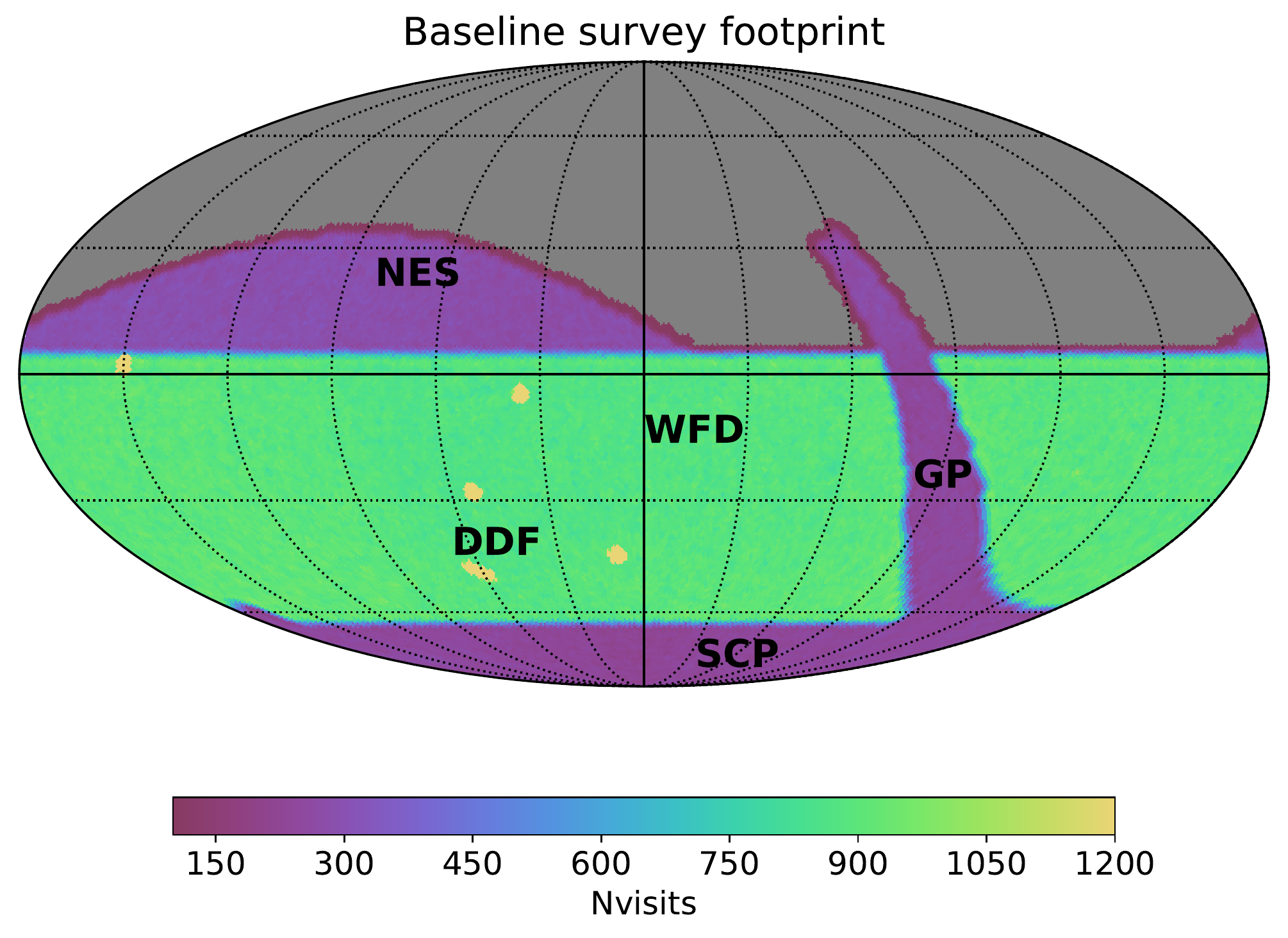}
\includegraphics[width=0.45\textwidth]{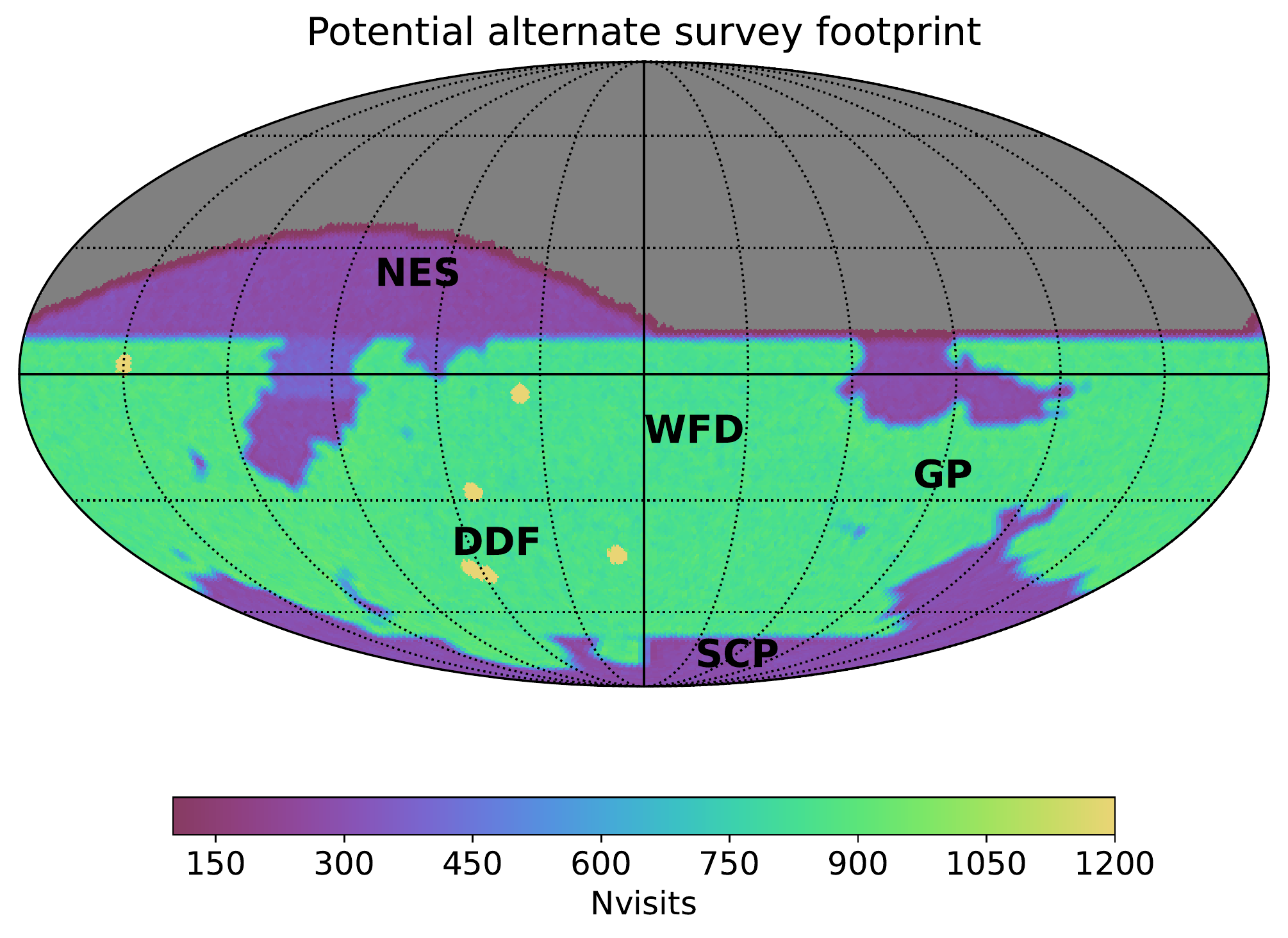}
\caption{LSST Footprint: the color encodes the number of visits across all filters for each position on the sky. The general survey footprint must meet the SRD design specifications: the WFD covers $\gtrsim18,000$ deg$^2$ to a median of 825 visits. Other surveys, like the Deep Drilling Fields (DDFs, which appear in yellow due to the higher number of observations), North Ecliptic Spur (NES), Galactic Plane (GP, which contains a large region on the bulge side, but also a smaller footprint on the anti-bulge side), and South Celestial Pole (SCP) that are shown in this figure, cover regions outside of the WFD footprint and have more flexibility in the observing strategy, including the number of visits and their distribution among filters. The initial survey footprint was envisioned as in the left panel 
(map generated from \opsim\ \texttt{baseline\_nexp2\_v1.7.1\_10yrs}, see \autoref{ss:software}), but details of the visit distribution for optimal science return are still being determined, including if this general survey footprint should be modified. 
Expansion of coverage into additional low dust extinction area and higher coverage of a portion of the Galactic plane, for example, may result in a footprint conceptually similar to what is shown on the right
(generated from \opsim\ \texttt{footprint\_6\_1.7.1\_10yrs}). This figure is discussed in \autoref{sec:intro}.}
\label{fig:footprint}
\end{figure*}

\begin{figure*}[t!]
\centering
\includegraphics[width=1.6\columnwidth]{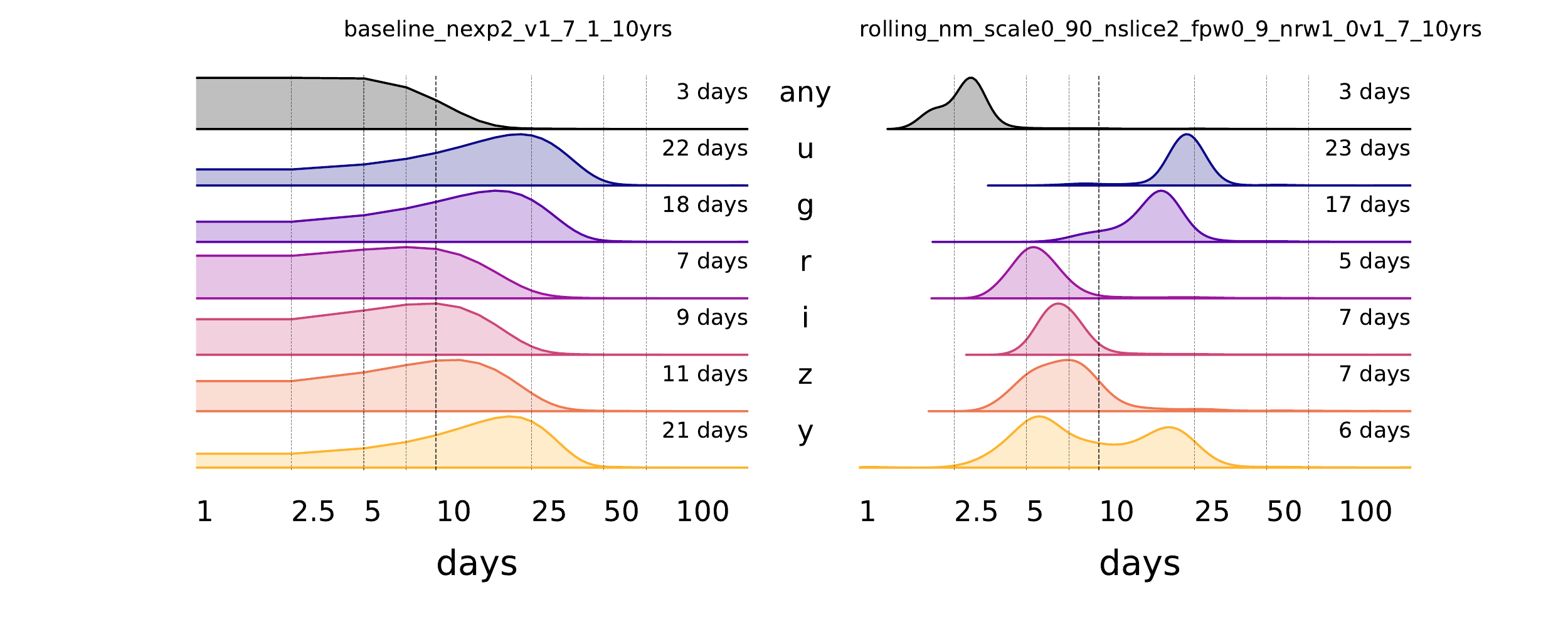}
        \caption{Distribution of median internight visit gaps (the time elapsed between visits to the same field in different nights) at a given location in the sky for two simulated LSST \texttt{OpSim} strategies: \texttt{baseline\_nexp2\_v1.7.1\_10yrs} (left) and \texttt{rolling\_nm\_scale0.90\_nslice2\_fpw0.9\_nrw1.0v1.7\_10yrs}  (right, for more details on the simulations see \autoref{ss:software}). Each curve shows the median internight gaps across pointing positions distributed on a \texttt{healpixel} grid with $N_{sides}=64$ (corresponding to $\sim0.85$~sq.~deg.). In the top row the distribution is shown for all observations, in the following rows for observations in each of the six Rubin LSST filters (as labeled). The distributions are smoothed via Kernel Density Estimation \citep{10.1214/aoms/1177728190, 10.1214/aoms/1177704472}; the location of the peak of the underlying distribution is indicated to the right of each curve. This figure shows that different \opsim\ runs, developed in compliance with the SRD requirements, can be very different even in core features. The figure is discussed in more detail in \autoref{sec:intro}. }
\label{fig:internightgap}
\end{figure*}

 A nominal survey footprint showing the distribution of visits across the sky, including WFD, some DDFs, and some potential minisurveys, is shown in \autoref{fig:footprint}. %It can be used as a benchmark the performance of all other survey proposals.
 Even a core parameter such as the median time elapsed between observations of the sky in different nights can be varied significantly for the WFD within the constraints of the SRD. 
\autoref{fig:internightgap} shows the distribution of inter-night gaps, a critical parameter to enable transient science, in two WFD survey simulations: a \texttt{baseline} proposal, \ie, a straightforward implementation of the survey design as described above, and a \texttt{rolling} cadence strategy, where the $\sim825$ visits for each sky pointing in the WFD are distributed unevenly over the 10-year LSST timeline, front-loading certain areas of the sky with higher density of observations early on, to later give way to intense observing of others. The rolling cadence is further discussed in \autoref{sec:whitepapers} and a more detailed description of the concept of rolling cadence and discussion of its trade-offs can be found in section 2.5 of \citealt{COSEP} and 4.9 of \citealt{jones2020} as well as inline at this URL\footnote{\url{https://project.lsst.org/meetings/ocw/sites/lsst.org.meetings.ocw/files/OpSim\%20Rolling\%20Cadence\%20Stratgey-ver1.3.pdf}}.

With LSST likely to start in 2024---a delayed re-baselined schedule due to the COVID-19 pandemic–––Rubin Observatory is undertaking the final 
planning for the initial observing strategy.
To ensure that the survey science potential is maximized and that the survey design does indeed serve the broad scientific community that will have access to the data, Rubin Observatory has, to an unprecedented degree, % for any precursor surveys, 
involved the scientific community in the survey design itself.  

To place in context publications 
containing science-driven proposals and recommendations for the survey strategy that are included in this Focus Issue, we describe here the LSST survey cadence optimization process. The process has been conducted openly and in close contact with the survey stakeholders and the scientific community at large.%, including %community 
%motivations and community feedback on operational decisions. 
 
The ongoing pre-operations phase of cadence optimization has included three major contributions by the science community: 
\begin{itemize}
    \item Early community engagement efforts, including the creation of software that enables the community to get involved in survey design. These are described in \autoref{ss:software} and \autoref{sec:early}. This phase culminated in the Community Observing Strategy Evaluation Paper \citep[][hereafter COSEP]{COSEP}, briefly described in \autoref{ss:COSEP}; 
    \item Cadence White Papers, solicited in 2018, briefly described in \autoref{sec:wp};
    \item Cadence Notes, solicited in 2020, and discussed in \autoref{sec:cn}.
\end{itemize}

%This Focus Issue collects community contribution to the optimization of the Rubin LSST observing cadence, produced in response to these solicitations. 
%, particularly a 2018 call for  White Paper in 2018 (\autoref{sec:wp}), and 2021 call for Cadence Notes (\autoref{sec:cn}). 
This Focus Issue provides the community with an opportunity to present the work that underlies many of these science-driven cadence recommendations.
 %The structure of this paper reflects the structure of the process itself: in \autoref{sec:early} we describe the initial phases of community engagement, including the software released by Rubin to facilitate the involvement of the scientific community in survey design. In \autoref{sec:wp} we describe the process leading to, and the results of, a 2018 call for White Papers on LSST Cadence. In \autoref{sec:cn} we describe the final phases of the optimization process which will lead to the initial survey design. Finally w
 We  highlight some lessons learned so far in this  community-focused process for survey design in \autoref{sec:lessonslearned} and \autoref{sec:ahead}.

\section{Open Software to enable community engagement}\label{ss:software}
To empower the community to make knowledgeable, science-driven recommendations, Rubin Observatory has released a series of simulated LSST pointing histories, using different survey strategies, and open-access software to generate quantitative analyses of these simulations.  These software tools originated as part of the overall Rubin Observatory simulations effort \citep{2014SPIE.9150E..14C}, and have evolved over time with considerable community input. 

\begin{figure}[t!]
\includegraphics[width=\columnwidth]{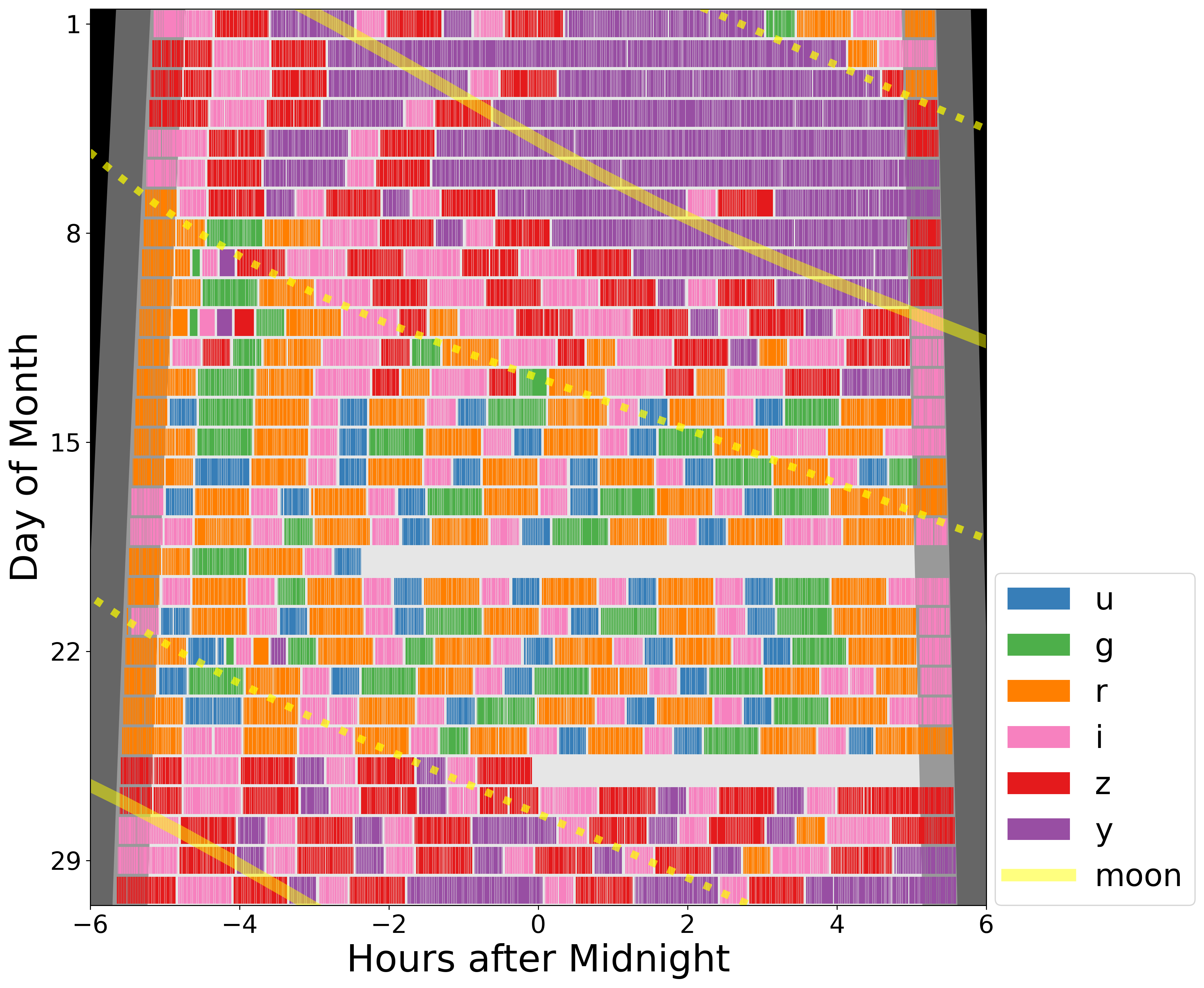}
\caption{A segment of an ``hourglass plot'' showing a sequence of observations in conjunction with astronomical constraints. A simulated calendar of observations for the month of April in year 1 of LSST operations is shown, color-coded by the bandpass of each visit (as indicated in the legend). The $x$-axis shows time away from midnight; the $y$-axis separates visits by the day of the month. The gray edges of the plot indicate twilight time, with observations continuing to $-12$ degree twilight in redder filters. Astronomical ($-18$ degree) twilight is shown with light gray, $-12$ degree twilight is dark gray, and periods where the Sun is above the horizon are in black. The time of lunar transit is shown as a wide yellow stripe; lunar rise and set times are shown by dotted yellow lines.  Redder filters are used when the Moon is up and bluer filters when the Moon is down and the sky is dark. Each 30-second visit is separated from adjacent visits by thin white vertical lines indicating the time between exposures, which depends on the time it takes the telescope to slew to a new pointing and it is typically short (and often not visible in practice here).  Thicker lines indicate longer than typical slew and setup times; for example, the time to change filters is about two minutes. Weather and telescope maintenance are also simulated: observing time lost to weather or telescope maintenance appears as larger white gaps on days 19 and 26. This figure is discussed in \autoref{ss:software}.}
\label{fig:hourglass}
\end{figure}

To generate potential realizations of the LSST surveys, Rubin developed a simulator that works with the LSST scheduler, collectively known colloquially as the Operations Simulator, or \opsim\ for short \citep{2019AJ....157..151N, 2016SPIE.9910E..13D, 2014SPIE.9150E..15D}, which can generate databases containing 10-year pointing histories, complete with weather, seeing, and sky brightness information \citep{2016SPIE.9910E..1AY}. A short sample of a resulting simulated observing history is shown in \autoref{fig:hourglass}. The version of the scheduler used at the time of writing, generally referred to  as the `Feature Based Scheduler' or FBS, is the fifth generation of scheduler codes developed for the LSST.\footnote{\url{https://community.lsst.org/t/from-opsim-v4-to-fbs-1-2/3856}}
In general, this software is used by Rubin Observatory to produce and release databases of observations following recommendations from the community.\footnote{\url{https://community.lsst.org/tag/opsim}} Sets of \opsim\ runs are released as ``families'' in which survey strategy parameters are varied along a specific simulation axis. For example: the \texttt{wfd\_scale} family explores allocating different fractions of the overall survey time to the WFD, the \texttt{footprint} family of simulations modifies the survey footprint according to different recommendations, and the \texttt{filter\_dist} family varies the distribution of visits in the survey between different filters.\footnote{The name of the \opsim\ runs are composed of a root indicating the \opsim\ family (\eg, \texttt{baseline} or \texttt{footprint}), the specific parameters of the run within the family, the \opsim\ release (\eg, \texttt{v1.7.1}), and the duration of the simulation (\eg, \texttt{10yrs}).}

The second open-access software package released by Rubin to facilitate community engagement in survey design is the Metric Analysis Framework  \citep[][hereafter \maf]{2014SPIE.9149E..0BJ}, which is intended to enable the calculation of metrics pertaining to the observations in a standardized and easily extensible manner. The \maf~provides a simple API for calculating image properties (\eg, seeing, proper motion) over a variety of spatial scales as well as tools to predict the properties of photometric measurements for objects implanted in the survey simulations (including light curves). The \maf\ thus allows the user to create metrics associated with specific science goals, and to evaluate them over existing survey simulations (the \opsim\ runs). That is: the user can test scientific cases and identify effective (or ineffective) cadences associated with them. Along with the API, many ready-to-run metrics, generally referred to as \maf-metrics (hereafter simply metrics), have been released. Furthermore, the most advanced metrics have been the result of collaborations with or direct contributions from the community\footnote{\url{https://github.com/LSST-nonproject/sims_maf_contrib}} in the three phases of survey evaluation described below. The outputs of standard metrics on existing \opsim\ runs are provided to Rubin Observatory and made available online to the community to help guide survey strategy choices.\footnote{The outputs of survey metrics are available at \url{http://astro-lsst-01.astro.washington.edu:8081}}

\section{Community Engagement in LSST Cadence Optimization -- pre-2018}\label{sec:early}
 
Planning for LSST has been undertaken hand-in-hand with the community from the start of the project. For example, the choice of four main science themes was guided by the community-wide input assembled in the report of the Science Working Group of the LSST in 2004.\footnote{Science Working Group of the LSST \& Strauss, M. A. 2004, Towards a Design Reference Mission for the Large Synoptic Survey Telescope, \url{https://ls.st/Doc-26952}} Similarly, the current design of the  DDFs was driven by a set of eight science white papers\footnote{Available from \url{https://www.lsst.org/scientists/survey-design/ddf}} written in 2011 by about 75 members of the LSST Deep-Drilling Interest Group. This led to the current selection of four DDF fields that maximize synergy with legacy data and ongoing surveys to enable a number of extragalactic science investigations.\footnote{\url{https://www.lsst.org/scientists/survey-design/ddf}} The Rubin Observatory Science Advisory Committee\footnote{\url{https://project.lsst.org/groups/sac/welcome}}  (SAC) has been providing valuable advice to Rubin on cadence issues since its inception in 2014.

\subsection{The LSST Scientific Community and the Science Collaborations}

Rubin Observatory and its construction and operation teams include 100's of people supported by investments of the NSF and DOE\footnote{Financial support for Rubin Observatory comes from the National Science Foundation (NSF) through Cooperative Agreement No.\ 1258333, the Department of Energy (DOE) Office of Science under Contract No.\ DE-AC02-76SF00515, and private funding raised by the LSST Corporation.} to create the Observatory, and design and run the LSST, including preparing, distributing and supporting the usage of data and data products from the survey. Rubin’s plan to maximize science from the LSST includes crucial support, engagement, and coordination with the scientific community. The community, in turn, needs funding support from all possible sources, including the US agencies and philanthropic organizations, to produce “User Generated Data Products” that drive the science analyses to the fullest extent embraced by Rubin’s scientific vision and mission. User Generated Data Products come from the community and take the Prompt and Data Release Products produced by Rubin to the next level to ensure the mission of Rubin Observatory is fulfilled. These data products and analyses can only happen through a high level of planning, coordination, and effort from the scientific community. 

Embracing these exciting challenges and promises of the LSST,
the scientific community has organized into Science Collaborations (SCs).
The SCs were formed to provide a forum for
the community to interact with Rubin Observatory's construction team (then the LSST Project), and to make the scientific case to be
presented to the 2010 Decadal Survey \citep{lsstSB}. Today, the SCs are eight independent teams, self-governed and self-managed, that include  over 1000 scientists
from six continents\footnote{\url{https://www.lsstcorporation.org/science-collaborations}} \citep{bianco2019better}. They work in close contact with Rubin to prepare to turn the LSST data into science and to help the Observatory make scientifically informed choices. 
Thus the SCs have been optimally placed to be core contributors of cadence recommendations throughout the cadence optimization process. For example, early survey strategy investigations using some of the initial \opsim\ simulations were driven by SCs, including Dark Energy Science Collaboration (DESC) analyses of potential dither patterns \citep{Carroll14,Awan16} that motivated a random nightly shift of pointing centers to become the Feature Based Scheduler (FBS) default.
Nonetheless, Rubin has sought input on cadence optimization from the entire scientific community, and suggestions have been welcomed from any group, both within and outside of the community of data rights holders.

\begin{figure*}[!ht]
\centering

  \includegraphics[width=0.55\textwidth]{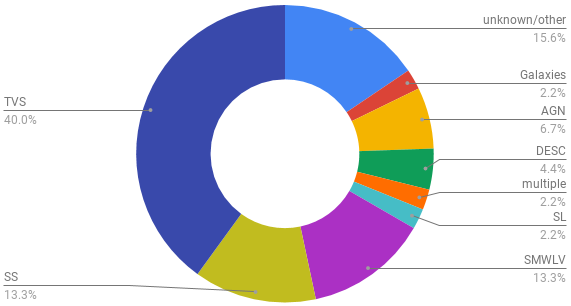}\includegraphics[width=0.3\textwidth]{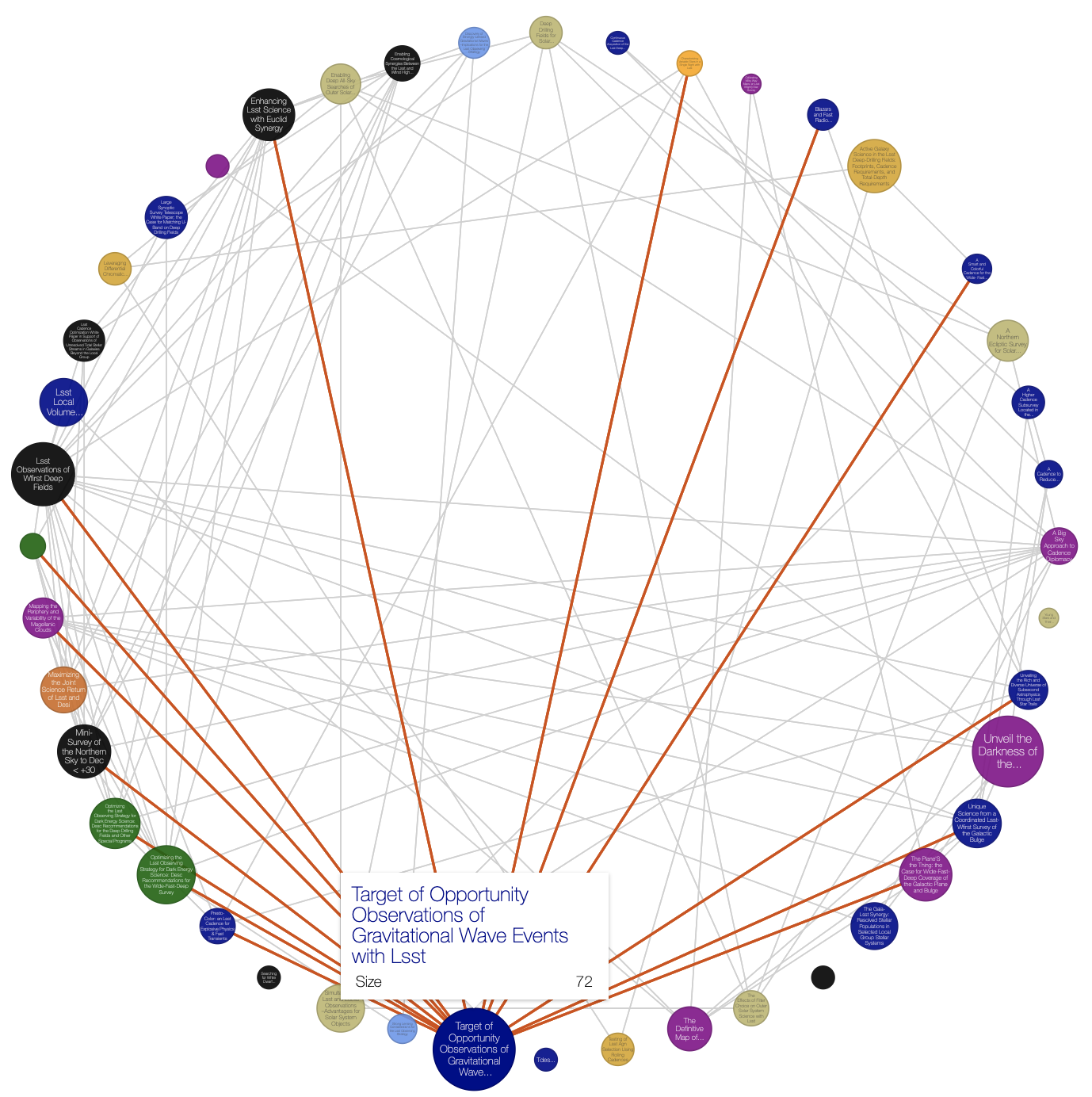}
\begin{minipage}{\textwidth}
\caption{A synopsis of the 46 Cadence White Papers submitted in response to the call issued in 2018: on the left the papers are grouped by the Rubin LSST Science Collaboration (SC) of the lead author with fractions from each SC indicated in the figure.  15\% of the papers were submitted by lead-authors unaffiliated with a SC.  %In the left panel the fraction of papers submitted by a SC according to the membership of the paper’s lead author is indicated. 
The right panel shows a screenshot of an interactive plot (available at \url{http://fbb.space/LSSTWP/WPbyfirstAuthor.html}) describing the co-authors network through a chord diagram. Each circle represents a paper, the size of the circle reflecting the size of the authoring team (the largest is 72 co authors for Target of Opportunity Observations of Gravitational Wave Events with LSST), color-coded, as in the left panel, by the SC of the lead author. Papers are linked by their co-authors. This figure is discussed in \autoref{sec:wp} and  \autoref{sec:lessonslearned}. }
\label{fig:whitepapers}
\end{minipage}
\end{figure*}

\subsection{Community Observing Strategy Evaluation Paper}
\label{ss:COSEP}

Members of the LSST science community gathered in 2015 to help design an observing strategy that would maximize the scientific output of the survey. This community includes scientists primarily (but not exclusively) drawn from people engaged in the LSST SCs and LSST construction Project. %They aim to provide quantitative feedback about how any proposed observing strategy would impact the performance of LSST science cases and thus enable robust decisions to be made when the telescope schedule is eventually set up. 

The core product delivered by this group is the \citetalias{COSEP}, a document co-authored by over 100 scientists and currently over 300 pages long, that lists a compendium of ideas and results. It is designed as a living document with the aim to bring together the group of people who are thinking about the LSST observing strategy problem, and facilitate their collective discussion.\footnote{The \texttt{GitHub} repository containing the living source for the \citetalias{COSEP} is {\url{https://github.com/LSSTScienceCollaborations/ObservingStrategy}}} It provided a first venue for evaluations produced by the community through the systematic use of the \maf and enabled the evaluation of the survey throughput based on a set of simulations representing small variations of the baseline survey for a variety of science cases.  %(although, due largely to the novelty of the software, not all science cases were fully integrated in the \maf framework, see \autoref{sec:lessonslearned}).  %Its audience is the entire LSST science community. %and most notable its SAC and Project Scientists who together will in the end decide what the LSST observing strategy will be. 
%This white paper is a vehicle for the community to communicate to the LSST Project, while the baseline observing strategy continues to be improved.
In order to standardize various constraints derived from diverse science cases and enable comparisons, each section contains the explicit answers to  ten questions  (available in full in \autoref{appendix}) designed to examine constraints and trade-offs between, for example: sky coverage and depth; uniformity and frequency of sampling (\eg, a rolling cadence);  single-visit depth and number of visits; Galactic Plane coverage (spatial, temporal, or depth); DDF sampling and depth; fraction of observing time allocated to each band; pairing of filters; and any requirements placed on commissioning.

The COSEP is articulated in nine topical chapters that cover 25 specific science cases. Additional sections discuss minisurveys (\eg,  covering specific sky locations and proposing modified exposure parameters), and plans for  coordinating LSST observations with other missions (\eg, the Nancy Grace Roman Observatory, \citealt{2015arXiv150303757S}, then WFIRST).

%A total of 76 detailed answers for 20 major science cases provided actionable input. 
The SAC reviewed this work and made recommendations to Rubin that shaped the next phase of the survey strategy optimization, including: %work and shared a detailed summary of the lessons derived from it.%\footnote{\url{https://project.lsst.org/groups/sac/sites/lsst.org.groups.sac/files/OpSim_experiments.pdf}}. The main recommendations include:
\begin{enumerate}
\item The Rubin construction Project should implement, analyze, and optimize the rolling cadence idea (driven by supernovae, asteroids, short timescale variability).
\item The Rubin construction Project should execute a systematic effort to further improve the ultimate LSST cadence strategy, including optimization of the sky coverage, $u$-band depth, filter pairing, minisurveys, DDFs, etc.
\end{enumerate}

\section{The 2018 Cadence White Papers solicitation}\label{sec:wp}

In part as a result of the SAC recommendations, and with an overarching goal of maximizing the science impact of the LSST, in 2018 Rubin Observatory issued a call to the entire scientific community for white papers on survey strategy ideas. Although the COSEP is a living document that can continue to be updated, this call
expanded the reach of the community engagement in the survey optimization process. The call\footnote{The call for white papers is available at \url{ http://ls.st/doc-28382}} solicited discussions of science cases in which Rubin could have an impact and associated science-driven cadence suggestions for the main survey, the minisurveys, the DDFs, as well as the first Target of Opportunity strategy suggestions for Gravitational Wave counterpart detection with Rubin \citep{2018arXiv181204051M}. It was supported by a new and more extended set of \opsim\ runs, itself enabled by progress in the simulator development\footnote{Changes to \opsim\ leading to the White Papers Call are described in \url{http://ls.st/Document-28453.}} (\autoref{ss:software}). 

The aim of the Cadence White Papers call was to create a portfolio of survey ideas, to be vetted and prioritized by the SAC, that would become the basis for a larger and more comprehensive set of \opsim\ runs. %, such as suggestions on the sky coverage, observing time allocation per filter, temporal coverage, and various detailed observing constraints. 
The 46 submitted white papers\footnote{\url{https://www.lsst.org/submitted-whitepaper-2018}} represent a wide swath of the astronomical community,
and, together with the \citetalias{COSEP}, shaped the next stage of the survey strategy evaluation.  Most submissions arose within SCs (\autoref{fig:whitepapers}, left) and many are the result of collaborative work across SCs, as demonstrated by the inter-connectivity of the authors' network (\autoref{fig:whitepapers}, right), but contributions were submitted as well by authors outside the SCs and interest groups related to other surveys (\eg, the Nancy Grace Roman Observatory, then WFIRST, \citealt{2015arXiv150303757S} and Euclid, \citealt{capak2019enhancing}).
The contents of these white papers were distilled into several areas for investigation by the SAC in their 2019 report.\footnote{The SAC report on the White Papers is available at \url{http://ls.st/doc-32816}}
%Distilling 46 detailed white papers into these actionable recommendations was a major contribution by the SAC to the LSST cadence optimization process. 
The ongoing cadence optimization process discussed in detail below is a direct result of those recommendations. The SAC also identified some vulnerabilities in the process: ``many of the white papers only outline what an appropriate metric would be, and coding these up will require substantial effort.  The SAC is concerned about how this work will be done;[...]  we recommend that the \opsim\ team be given the resources to code the metrics suggested in the white papers [...]''.

\subsection{Survey Simulations and Cadence Optimization following the White Papers}\label{sec:whitepapers}
 
Following the SAC response to the 2018 White Papers, the Rubin survey strategy team produced a series of simulations exploring the particular survey strategy options recommended for further investigation. These include investigations into the survey footprint, the amount of survey time devoted to WFD observations, various options for pairing visits within a night or adding more nightly visits, the exposure time per visit, enforcing seeing requirements at each point in the sky in certain filters, dithering options for the DDFs, and exploring the effect of adding specific minisurveys, such as short exposures, twilight Near Earth Object discovery visits, or additional visits at high airmass for better Differential Chromatic Refraction measurements,  and more. 

Sets of these new simulations were released in groups. For this call,  a total of 173 \opsim\ runs were made available under \opsim~releases 1.5 to 1.7.1. %released for the Cadence Notes. %; the bulk of these were released in v1.5 but early community feedback led to additional simulations being produced in v1.7 and v1.7.1. 
These simulations and their analysis with sets of standard metrics are described in detail in  \citet[][hereafter LSST document PSTN-051]{jones2020}.\footnote{PSTN-051 is available at \url{https://pstn-051.lsst.io}.} 

\begin{figure}[t!]
\centering
\includegraphics[width=0.9\columnwidth]{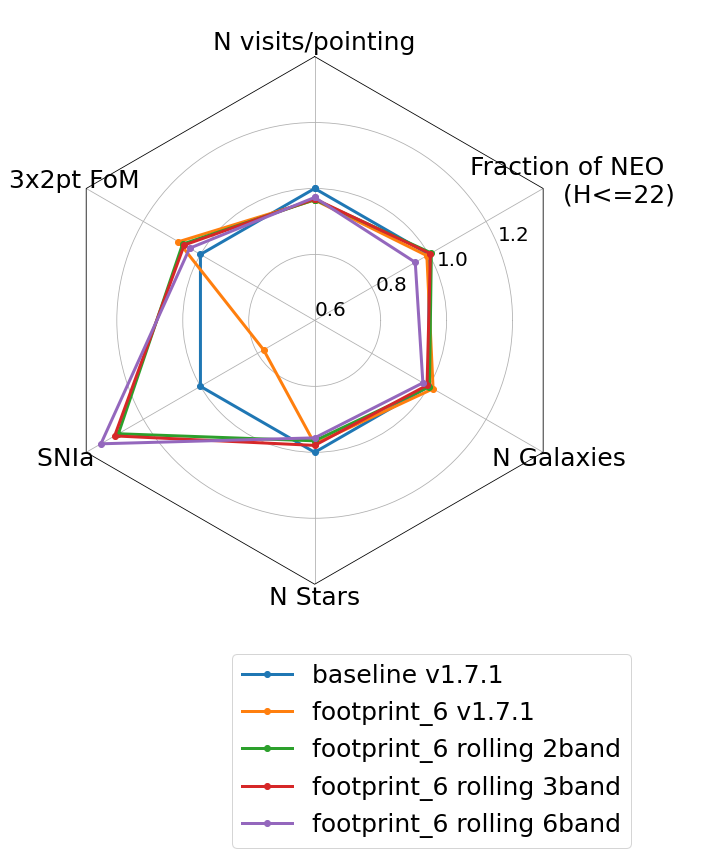}
\caption{Illustration of a ``radar plot'' comparing science metrics from a variety of core science areas for five \opsim\ realizations (as indicated in the legend). Each corner of the hexagon represents a different metric, whose value is mapped to the distance from the center of the hexagon. Clockwise from the top, the metrics measure: the median number of visits per pointing in the WFD (\texttt{N visits/pointing}), the fraction of Near Earth Objects of magnitude $H\leq22$ detected  (\texttt{Fraction of NEO}), the numbers of galaxies (\texttt{N Galaxies}) and stars (\texttt{N stars}) expected to be recovered, the expected number of supernovae type Ia (\texttt{SNIa}) observed sufficiently well to contribute to cosmological measurements \citep{lochner2018optimizing}, and a collective metric for static probes of the extra-galactic sky (\texttt{3x2pt FoM}, combining  Weak Lensing and Galaxy Clustering, \citealt{lochner2018optimizing}). This set of \opsim\ runs varies the footprint (which impacts the science throughput for cosmological supernovae as measured by \texttt{SNIa}) and introduces a rolling cadence (which improves \texttt{SNIa} performance as a denser cadence enables better characterizations at relevant, short time scales and the distribution of visits in a more concentrated sky area leads to small efficiency improvements). This figure is discussed in \autoref{sec:whitepapers}.}
\label{fig:radar_ex}
\end{figure}

%Detailed metrics, written through the \maf\ API, to analyze the science return of these simulations have been created by Rubin survey strategy team and by the community, and more continue to be developed. 
The metrics written to analyze the science return of these simulations through the \maf\ API, both those created by the Rubin survey strategy team and those created by the community, need to be ultimately combined into a comprehensive view of an \opsim\ run that can enable comparative studies and, in the end, the definition of the survey design. This is a complex exercise in and of itself, which is at the heart of the last phase of the survey cadence optimization process described in \autoref{sec:cn}. An example of a visualization of a subset of metrics indicating the effect of rolling cadence on a variety of science areas is shown in~\autoref{fig:radar_ex}.

\section{The 2020-2022 Optimization Phase}\label{sec:cn}
  
With the publication of \citetalias{jones2020}, Rubin (including now both the construction and the early operations teams) has started the final phase of the LSST cadence optimization before the start of the 10-year survey with the creation of the Survey Cadence Optimization Committee. 

\subsection{The Survey Cadence Optimization Committee}\label{ss:soc}

The Survey Cadence Optimization Committee (SCOC) is an advisory body to the Rubin Observatory 
Operations Director . The committee was formed in 2020, and it will be a standing committee throughout the life of Rubin Observatory operations. The SCOC is responsible for optimizing the LSST cadence within the constraints imposed by the observing system, observing conditions, 
science drivers, and scientists invested in its mission and legacy. Its tasks are as follows:
\begin{itemize}
    \item Make specific recommendations for the initial survey strategy for the full 10-year survey and disseminate these recommendations via public reports and on-going engagement with the community.
    \item Make specific recommendations for “Early Science” observations, which might be carried out after the end of commissioning during the first months of operation of Rubin Observatory.
    \item Continue its activities throughout Operations by evaluating reports prepared by the Survey Evaluation Working Group, a project-internal group set up to measure the performance of the survey and scheduler, and make necessary recommendations for adjustments of the survey strategy.
\end{itemize}

The SCOC consists of ten voting members (12 total)\footnote{See \url{https://www.lsst.org/content/charge-survey-cadence-optimization-committee-scoc}}, named by the SAC. Most members are drawn from the science community and are not Rubin Observatory employees. %It is chaired, however, by the Rubin Observatory LSST Head of Science and includes the Rubin Observatory System Performance Scientist, as non-voting members. 
%While the SCOC members are largely selected from the existing LSST SCs, they do not represent the specific scientific interests of an LSST SC, but (like the SAC) work to optimize the global scientific productivity of the Rubin Observatory LSST.
The SCOC membership is expert and diverse, and its deliberations are transparent and inclusive.\footnote{For the SCOC charge, membership, meeting minutes, recommendations and other documents, please see \url{https://ls.st/scoc}} %\question{should we mention the liaisons? they are important as through the liaisons some SCs have asked extensions of the simulations. }
To ensure continuous communication with the scientific community, the SCOC members also serve as designated liaisons with each SC. This direct line of communication between the SCs and SCOC allows the SCs to be kept informed about the ongoing simulation generation process, and promptly identify gaps in the landscape of simulations or issues with the implementation of metrics that measure their efficiency (see also \autoref{sec:lessonslearned}).

The SCOC cadence recommendation process is itself split into two phases. The first
phase, to be concluded by the end of calendar year 2021, aims to select a
cadence family that will capture an overall strategy. By the end of calendar year 2022, the SCOC will fine-tune the selected cadences to come to a final recommendation for the initial cadence to begin execution when operations starts, though this recommendation will be continuously re-evaluated and revised if necessary after the start of operations, as discussed in \autoref{sec:intro}.  %After several iterations, primarily driven by COVID-related 
%schedule updates at the Project level, 
%SCOC has adopted the following schedule for 
%its work: 
%\begin{itemize}
%    \item Aug 2020: publication of the PSTN-051 report
%    \item Nov 2020: the 1$^{st}$ cadence workshop\footnote{See https://project.lsst.org/meetings/scoc-sc-workshop/about}
%    \item Apr 2021: the Cadence Notes submission deadline 
%    \item Aug 2021: SCOC publishes draft ``phase 1" recommendation 
%    \item Nov 2021: the 2$^{nd}$ cadence workshop
%    \item Dec 2021: finalized SCOC recommendation for ``phase 1" 
%    \item Mar 2022: simulations of the recommended strategy become available 
%    \item Aug 2022: tentative 3$^{rd}$ workshop to fine-tune the recommended   
%           strategy, including "early science optimization" 
%    \item Dec 2022: the simulation of the adopted observing strategy (the new 
%          LSST baseline) produced and made publicly available
%    \item Apr 2023: the observing strategy fixed and implemented in the Scheduler 
%           and the Observatory Control Software 
%\end{itemize}
A detailed schedule of the SCOC activities and decision process, including workshops that put the SCOC in direct contact with the scientific community at large,  can be found on the SCOC web pages.\footnote{\url{https://www.lsst.org/sites/default/files/SCOC\%20Handout.pdf}}

\subsection{Cadence Notes}\label{ss:cn}

%There are over 100 simulated cadences described in \citealias{PSTN-051}.
The COSEP and the cadence white papers provided the necessary framework to identify needed simulations, as well as defining how further cadence recommendations should be structured to be maximally useful in the optimization process. To receive formal feedback from the SCs and other 
stakeholders about this new generation of simulated surveys, and with a more significant adoption of the \maf\ software by the community, the SCOC issued
one further call\footnote{ \url{https://ls.st/cadencenotes}} requesting community input, in the form of Cadence Notes, to explore and evaluate survey strategy options presented in the \citetalias{jones2020}. Seven specific questions (listed in full in \autoref{appendix:cn}) were posed in the Cadence Notes.

Following the SAC recommendations, in this stage of the process the Rubin LSST survey strategy team was in close communication with the SCs and the scientists working on Cadence Notes, offering, for example, regular office hours. The SCs self organized to collaboratively work notes and share expertise on the use of the \maf software.\footnote{\eg, \url{https://lsst-tvssc.github.io/metricshackathon2020/}}%These questions aim at assessing for each science case the potential advantages of extending the WFD footprint beyond 18,000 sq.deg., of employing a rolling cadence, of extending exposure length in the $u$ band,  recommendations for minisurveys and DDFs,  for balancing the time allocated for each band, for the pairing of filters, and recommendations or constraints on dithering patterns. 

The community submitted 39 Cadence Notes.\footnote{Available from \url{https://www.lsst.org/content/survey-cadence-notes-2021}} These Notes are under review by the SCOC and they represent a
significant additional source of information for the SCOC in its deliberations
towards the ``phase 1" recommendation. 
The definition of the initial Rubin LSST observing strategy is yet to be finalized at the time of this writing. The remaining phases of the survey optimization process will be described in detail in the closing paper of this Focus Issue.

\section{Lessons Learned So Far}\label{sec:lessonslearned}

Over the last six years of intensive community involvement in optimizing LSST's observing strategy, many initial lessons have been learned.  One key lesson has been that effective involvement of the broad community in LSST's survey strategy requires significant coordination across all elements of the Rubin ecosystem, including the construction Project, SAC, and SCs. 

%As described above, that coordination has required multiple steps.  
A multi-step process was required to (i) engage the most diverse possible user community, (ii) to iterate on initial community analysis and construction Project progress, and (iii) to be agile in response to evolving scientific opportunities. This process began with pre-construction community engagement on the choice of LSST DDFs. Since then, survey input from the community has been solicited in three phases: 

\begin{itemize}
    \item through the creation of a single collaborative document, the \citetalias{COSEP}, employing, for the first time, community analysis through the \maf and \opsim;
    \item through the submission of White Papers broadly including new cadence ideas and considerations; 
    \item and finally through the submission of Cadence Notes to specifically review a comprehensive set of \opsim\ runs, and fine-tune survey parameters that remain open to optimization.
\end{itemize}  

\autoref{tab:calls} summarizes the key differences between these three phases.  We highlight five aspects of our community engagement process that have been important to its success to date:

%\caption{Community contributions to Rubin LSST cadence design}

%I think itd be nice to show a table, with a column for each call for input, highlighting the differences between those calls. Rows in the table could be: date; goal of call; information needed; tools available; target audience;  response level. What do you think?

%\end{table*}

\begin{table*}
\renewcommand{\arraystretch}{1.5}
\centering

\begin{tabular}{ >{\raggedleft\arraybackslash}p{2cm}  |  p{4.1cm} | p{4.1cm} | p{4cm}   }  
\tablewidth{0pt}
& \textbf{COSEP} & \textbf{White Papers} & \textbf{Cadence Notes}\\
\hline  
\textbf{date} & 2015---2017 & 2018, June (call) --- 2018, November (submission deadline) & 2020, December (call) --- 2021, April (submission deadline) \\
\textbf{goal} & To explore the effects of changes to the baseline survey strategy as specified by the SRD on
the detailed performance of the anticipated science investigations &  To propose significant modifications of the survey strategy, including minisurveys & 
To evaluate a broad collection of \opsim\ runs that implement suggestions from the white papers with refined metrics\\

\textbf{target community} & The LSST Science Collaborations & Open submission & Open submission \\ % Lynne: "The LSST Observing Strategy community" could be just "The LSST community" because the COSEP was pretty widely sourced from the community Phil: How about "LSST Science Collaborations"? I think that is who we mainly targeted. And the COSEP lives in teh SCs' GitHub

\textbf{response} & Single document articulated in 9 topical chapters, {25} science cases, three suggestions for minisurveys, and a discussion of synergy with space-based surveys (Roman Observatory, then WFIRST), 104 unique authors & 46 papers, 467 unique authors & 39 notes, 218 unique authors\\

\textbf{format} & Open discussion and 10 questions~(\autoref{appendixCOSEP}) &
Open discussion, ranked-priority table, and 5 questions~(\autoref{appendixwp}) & 7 questions~(\autoref{appendix:cn})\\

\textbf{\opsim\ version} & v3.3.5 & v4 & FBS\\

\textbf{number of available \opsim\ runs} & 14 & 16 & 173\\

\textbf{\opsim\  software changes} & & Optimizations in the rate of observing over time to better balance the time coverage of the minisurveys over the survey lifetime, ability to strongly prioritize observing closer to the meridian. & Modularized survey algorithms to allow for more flexible reward calculations (beyond just slew-time and target maps),  leading to more flexible observing strategies. Evolutions of the scheduler algorithm between simulation releases 1.5 and 1.7.1 allowed improved rolling cadence simulations, where visits are distributed on variable timescales in regions of sky over the 10-year survey. \\

\hline
\end{tabular}

\caption{Community contributions to Rubin LSST survey strategy design}\label{tab:calls}

\end{table*}

{\bf 1. Simulated Surveys and Open Analysis Software:} - The release of open software that supports direct interaction with the LSST simulations (\opsim~and \maf, see \autoref{ss:software}) was critical to enable contributions from the community. 
Rather than simply soliciting suggestions on preferred strategies, which would lead to a large set of options without a clear path to optimization, this software framework allowed the users to interact with the simulated survey and see the impact of cadence changes on a science case in conjunction with astronomical and technical constraints, as well as the impact that the same changes have on other science cases. The preparation and release of these software packages were key steps in the process of involving the community in survey strategy decisions.

The provision of software environments and support for using them was essential for including a wide range of astronomers in \opsim\ analysis and \maf-metrics development.
In the White Paper phase, Docker images \citep{merkel2014docker} were provided so that astronomers did not need to install and maintain the full set of interdependent software packages required to use \maf.
In later phases, virtual analysis platforms such as the SciServer\footnote{\url{https://www.sciserver.org/about/}} and NOIRLab's Astro Data Lab\footnote{\url{https://datalab.noirlab.edu/}} provided accounts to scientists working on Rubin optimization.
The \opsim\ databases --- which are $ \sim 1$~GB each --- were made available on these platform alongside software environments, example codes, and compute resources for processing.
Use of these platforms lowered the barrier to participation and enabled collaboration among scientists.

{\bf 2. Leverage SC expertise} - Developing, maintaining, releasing and supporting software and training the community in its usage are expensive, time-consuming activities. %This kind of deep engagement with the scientific community at large in the construction phases of a project is unusual and demanding. 
Future surveys that wish to engage the community in the survey design process as Rubin did should scope resources specifically for these dissemination and support activities. %\question{FBB: I am trying to say that more resources should have been given for this by NSF/DOE..... but in a nice way.... did i make it?}. 
With limited resources available to the Rubin LSST survey strategy team, the SC environment became critical for sharing knowledge and know-how and bringing this software to a large community of users by employing a sort of self-arranged ``train-the-trainer'' model \citep{pearce2012most}.

In order to respond to the opportunity to contribute to the survey design, Rubin Observatory, the SCs, and the LSST Corporation\footnote{\url{https://www.lsstcorporation.org/}} (a non profit engaging in fund-raising to support the scientific community in preparation for LSST) have organized a number of workshops, hackathons, and meetings to enable knowledge transfer, exchange of expert opinions, and communication with the Observatory. The Rubin survey strategy team has participated in many of these gatherings making suggestions on metrics and metric implementation. 
Some of these activities were hosted within Rubin meetings (\emph{e.g.}, at the annual Project Community Workshop, the all-hands Rubin meeting), others within SC meetings, others yet in purposely organized events (\eg, the Heising-Simons Foundation sponsored a Cadence Hackathon supporting the participation of members of all SCs with travel grants and the development of Cadence White Papers with seed grants\footnote{\url{ls.st/hcca}}). 
The SCs particularly acknowledge the support of the LSST Corporation in securing and directing private funds toward activities that supported the recommendations presented to Rubin.

The SCs have been leveraging their internal network structure to coordinate the presentation of different scientific cases within and across SCs. Being in close communication with Rubin allowed the SCs to be central in this process and contribute scientific expertise and insight (see also \autoref{sec:early} and \autoref{sec:cn}). The majority of the 2018 White Papers and 2021 Cadence Notes were created within or in connection with one or more SCs (\autoref{fig:whitepapers}). However, this exercise stretched the organizational capabilities of an un-funded organization that operates largely on a volunteer basis \citep{bianco2019better}. 

Some SCs converged on a single document to present their recommendations with one voice (\eg, Dark Energy SC –DESC–, Solar System SC –SSSC–, Strong Lensing SC -SLSC-).  Other SCs (\eg, the Transient and Variable Stars SC –TVS–, that covers topics related to Time Domain Astronomy in both the Galactic and Extragalactic environments, Stars, Milky Way, and Local Volume SC –SMWLV–, AGN SC) have responded to the calls for contributions with multiple papers and notes.  In practice, the ability to present a single response, which requires internal reconciliation of competing science cases, has depended on the breadth of the science encapsulated within a single SC, as those with a greater diversity of science cases require more substantial management effort to enable that reconciliation. Coordination between SCs with overlapping interests also required additional work: for example, SMWLV and TVS collaborated closely on Galactic science topics and produced additional documents to the SCOC to summarize the SC contributions that were split among multiple notes. The SCs,  except for DESC,\footnote{DESC efforts on observing strategy are partially supported under DOE Contract DE-AC02-76SF00515} have no operational budget to support these efforts (see also \autoref{sec:ahead} and  \citealt{bianco2019better}).

{\bf 3. Providing Templates} - Expecting a large response, Rubin provided templates for the White Papers and Cadence Notes that ensured the responses would be concise, and limits to the number of pages were provided. This framework was helpful to limit the amount of work required of the committees reviewing the papers, and of the community itself. This Focus Issue, then, provides the opportunity to share the detailed work that underlies these papers and notes.

{\bf 4. Reliable and Continuous Communication} - Ensuring continuous communication with the entire scientific community was also crucial in this process: for example, new \opsim\ runs need to be distributed broadly and promptly for the community to be able to answer the most relevant questions regarding survey strategy contributions. In addition to providing direct lines of communication with the SCs (\autoref{ss:soc}), Rubin kept the community at large updated on discussion about new simulations and metrics or cadence-related events via the Rubin Community public  forum\footnote{See for example \url{https://community.lsst.org/t/july-2019-update/3760}}, while metrics contributed by the community and vetted by the Rubin survey strategy team are collected on a dedicated \texttt{GitHub} repository.\footnote{\url{https://github.com/LSST-nonproject/sims_maf_contrib}} As the process of cadence optimization moves through its final stages, more simulations will be produced and shared to respond to particular questions and aid in tuning the chosen survey strategy. Even during Rubin operations, additional simulations will be prepared for yearly evaluations of the ongoing survey and to determine best choices of survey strategy for the remainder of the LSST.

{\bf 5. Aiming For Broad Community Input} - Rubin's community approach to survey optimization was designed with the goal to enable all potential users to have a voice in  LSST survey strategy planning.  Scientists from a wide range of institutions participated in the process. The SCs, for example, include members affiliated with research-focused institutions (R1), teaching-focused institutions, Community Colleges, Labs, and virtual institutes. The geographic span of the SCs reaches five continents. 

Enabling the publication of the work underlying the cadence recommendations in peer-reviewed journals is an important step to ensure that individuals outside of the Rubin Observatories employees group can get academic recognition for the time they invest in the process. This Focus Issue collects work related to the optimization of the Rubin LSST observing strategy, as produced by the scientific community throughout the multiple phases of the cadence optimization, providing an opportunity to present the work that underlies 2021 Cadence Notes presented to Rubin Observatory, but also including earlier work, such as the 2018 White Papers.

\section{A Look Ahead}\label{sec:ahead}

Although our community-driven optimization of Rubin's LSST cadence has delivered initial lessons and the science analyses described in this ApJS Focus Issue, challenges remain.  Community engagement in optimizing Rubin's WFD survey has increased Rubin LSST science potential and supported community preparation to conduct science, but it has also been a time and labor intensive endeavor.  It has taken six years of active community engagement in survey optimization to reach this point of our work.  We aim for this Focus Issue to support future survey-driven missions in their aim to maximize survey science through community engagement in survey optimization.

A significant challenge is that any community involvement plan such as this one remains selective unless participation support is provided.  The learning curve to master the cadence analysis remained significant through all three engagement phases. In many cases, this work did not fall under the specification of research grants and was done as service, supported by funds with unusual flexibility (\eg, start-ups and sabbaticals), or even performed outside of working hours. It was difficult for most members of the community to dedicate the time required to conduct these activities at a meaningful level, because not everyone is in a position to dedicate substantial unfunded time and effort to this work owing to workload and funding constraints.  Very few are able to do so for extended periods of time. Likewise, the work of the committees in the evaluation of survey strategy input from the community is a time consuming activity.  This effort was also uncompensated and performed as service to the scientific community. This framework biases the contributions to the cadence and the evaluation committee membership toward certain job profiles and seniority levels, particularly toward well-funded scientists with significant job flexibility, potentially also biasing the domain expertise to well-funded areas of astrophysics.

The closing paper of this Focus Issue will detail the cadence choice planned for the start of Rubin LSST and the process and considerations that led to its finalization.  This opening paper has outlined the initial lessons learned from our community engagement to date. This is not the end of the story.  The optimization of the observing strategy in an on-going process with the goal of maximizing science through fine-tuning the LSST survey strategy.  In the months and years to come, we expect to learn new things about the most effective way to incorporate community input into this survey optimization. 

 %While these SCs have produced multiple papers and notes in response to R, these documents have been coordinated to identify synergies and tensions throughout.

\vspace{2cm}
{\bf Acknowledgments}

The authors wish to thank the Rubin internal reviewers, including Dr. Rahul Biswas, who helped improving the quality of the paper.

This material is based on work supported in part by the
National Science Foundation through Cooperative Agreement
1258333 managed by the Association of Universities for
Research in Astronomy (AURA), and the Department of
Energy under Contract No. DE-AC02-76SF00515 with the
SLAC National Accelerator Laboratory. 

Additional LSST
funding comes from private donations, grants to universities,
and in-kind support from LSSTC Institutional Members. The authors acknowledge the support of the LSST Corporation in securing and directing private funds toward activities that supported the community involvement with Rubin. 

The authors acknowledge the support of the Vera C. Rubin LSST Science Collaborations that provided a collaborative environment for Rubin related research and exchange of knowledge and ideas.

AJC acknowledges support from NSF award AST1715122 and DOE award DE-SC-0011635 

EG is supported by the US Department of Energy grant DE-SC0010008.

JS acknowledges support from the Packard Foundation.

MES was supported by  UK Science and Technology Facilities Council (STFC) Grant  ST/V000691/1.

ML acknowledges support from South African Radio Astronomy Observatory and the National Research Foundation (NRF) towards this research. Opinions expressed and conclusions arrived at, are those of the authors and are not necessarily to be attributed to the NRF.

RM is supported by the US Department of Energy grant DE-SC0010118.

SJS was supported  by UKRI STFC grants ST/S006109/1 and ST/N002520/1. 

TA acknowledges support from FONDECYT Regular 1190335 and Millennium Science Initiative ICN12\_009.

WNB thanks the V.M. Willaman Endowment at Penn State.

This document was prepared using resources of the Fermi National Accelerator Laboratory (Fermilab), a U.S. Department of Energy, Office of Science, HEP User Facility. Fermilab is managed by Fermi Research Alliance, LLC (FRA), acting under Contract No. DE-AC02-07CH11359.

This
research has made use of NASA’s Astrophysics Data System
Bibliographic Services.
% Eric Neilsen says we need this for Fermilab authors - not sure whereit goes in the list -- not sure what COLLABORATION NAME should be but perhaps Rubin Observatory? (although this is also not Rubin .. and also this sentence should be edited in properly above). 

\emph{Facility}: Rubin Observatory.

We used the following \texttt{Python} packages:
    \begin{itemize}
      \item \texttt{numpy} \citep{harris2020array}
      \item \texttt{maplotlib} \citep{matplotlib}
      \item \texttt{seaborn} \citep{seaborn}

\end{itemize}

%\end{acknowledgments} 

\appendix
\label{appendix}
\section{Guiding cadence contribution with targeted questions.}
To standardize the community cadence contributions, the \citetalias{COSEP}  (\autoref{sec:early}), White Papers (\autoref{sec:wp}), and Cadence Notes (\autoref{sec:cn}) templates included targeted questions aimed at identifying trade-offs and constraints for every science case and strategy suggestion offered by the community. A complete list of these questions follows here.
\subsection{Full list of questions included in the COSEP}\label{appendixCOSEP}

\begin{table}[b!]
    \centering
    \begin{tabular}{|l|l|l|l}
        \toprule
        Properties & Importance \hspace{.3in} \\
        \midrule
        Image quality &     \\
        Sky brightness &  \\
        Individual image depth &   \\
        Co-added image depth &   \\
        Number of exposures in a visit   &   \\
        Number of visits (in a night)  &   \\ 
        Total number of visits &   \\
        Time between visits (in a night) &  \\
        Time between visits (between nights)  &   \\
        Long-term gaps between visits & \\
        Other (please add other constraints as needed) & \\
        \toprule
    \end{tabular}
    \caption{{Constraint Rankings from Cadence White Paper template.} Summary of the relative importance of various survey strategy constraints. Please rank the importance of each of these considerations, from 1=very important, 2=somewhat important, 3=not important. If a given constraint depends on other parameters in the table, but these other parameters are not important in themselves, the authors were directed to only indicate the final constraint as important. For example, individual image depth depends on image quality, sky brightness, and number of exposures in a visit; if a science case depends on the individual image depth but not directly on the other parameters, individual image depth would be `1' and the other parameters could be marked as `3', giving maximum flexibility when determining the composition of a visit.}
        \label{tab:obs_constraints}
\end{table}

\begin{enumerate}
\item Does the science case place any constraints on the tradeoff between the sky coverage and
coadded depth? For example, should the sky coverage be maximized (to $\sim 30,000$ deg$^2$
, as e.g.,
in Pan-STARRS) or the number of detected galaxies (the current baseline of 18,000 deg$^2$)?

\item  Does the science case place any constraints on the trade-off between uniformity of sampling
and frequency of sampling? For example, a ``rolling cadence'' can provide enhanced sample
rates over a part of the survey or the entire survey for a designated time at the cost of reduced
sample rate the rest of the time (while maintaining the nominal total visit counts).

\item  Does the science case place any constraints on the tradeoff between the single-visit depth and
the number of visits (especially in the $u$-band where longer exposures would minimize the
impact of the readout noise)?
\item  Does the science case place any constraints on the Galactic plane coverage (spatial coverage,
temporal sampling, visits per band)?
\item Does the science case place any constraints on the fraction of observing time allocated to each
band?
\item  Does the science case place any constraints on the cadence for deep drilling fields?
\item  Assuming two visits per night, would the science case benefit if they are obtained in the same
band or not?
\item  Will the case science benefit from a special cadence prescription during commissioning or early
in the survey, such as: acquiring a full 10-year count of visits for a small area (either in all
the bands or in a selected set); a greatly enhanced cadence for a small area?
\item  Does the science case place any constraints on the sampling of observing conditions (e.g.,
seeing, dark sky, airmass), possibly as a function of band, etc.?
\item  Does the case have science drivers that would require real-time exposure time optimization
to obtain nearly constant single-visit limiting depth?

\end{enumerate}

\subsection{Full list of questions included in the White Paper templates}\label{appendixwp}

\begin{enumerate}
 \item What is the effect of a trade-off between your requested survey footprint (area) and requested co-added depth or number of visits?
    \item If not requesting a specific timing of visits, what is the effect of a trade-off between the uniformity of observations and the frequency of observations in time? e.g. a `rolling cadence' increases the frequency of visits during a short time period at the cost of fewer visits the rest of the time, making the overall sampling less uniform.
    \item What is the effect of a trade-off on the exposure time and number of visits (e.g. increasing the individual image depth but decreasing the overall number of visits)?
    \item What is the effect of a trade-off between uniformity in number of visits and co-added depth? Is there any benefit to real-time exposure time optimization to obtain nearly constant single-visit limiting depth?
    \item Are there any other potential trade-offs to consider when attempting to balance this proposal with others which may have similar but slightly different requests?
\end{enumerate}

In addition, the paper template contained a table for ranking constraints on the survey strategy, which we reproduce in \autoref{tab:obs_constraints}.

\subsection{Full list of questions to be answered in the Cadence Notes}\label{appendix:cn}

\begin{enumerate}
\item Are there any science drivers that would strongly argue for, or against, increasing the WFD footprint from 18,000 sq. deg. to 20,000 sq.deg.? Note that the resulting number of visits per pointing would drop by about 10\%. If available, please mention specific simulated cadences, and specific metrics, that support your answer. 

\item Assuming that current system performance estimates will hold up, we plan to utilize the additional observing time (which may be as much as 10\% of the survey observing time) for visits for the minisurveys and the DDFs (with an implicit assumption that the main WFD survey meeting SRD requirements will always be the first priority). What is the best scientific use of this time? If available, please mention specific simulated cadences, and specific metrics, that support your answer.

\item  Are there any science drivers that would strongly argue for, or against, the  proposal to change the u band exposure from 2x15 sec to 1x50 sec? If available, please mention specific simulated cadences, and specific metrics, that support your answer. 

\item  Are there any science drivers that would strongly argue for, or against, further changes in observing time allocation per band (e.g., skewed much more towards the blue or the red side of the spectrum)? If available, please mention specific simulated cadences, and specific metrics, that support your answer.

\item Are there any science drivers that would strongly argue for, or against, obtained two visits in a pair in the same (or different) filter? Or the benefits or drawbacks of dedicating a portion of each night to obtaining a third (triplet) visit?  If available, please mention specific simulated cadences, and specific metrics, that support your answer. 

\item Are there any science drivers that would strongly argue for, or against, the rolling cadence scenario? Or for or against varying the season length? Or for or against the AltSched N/S nightly pattern of visits? If available, please mention specific simulated cadences, and specific metrics, that support your answer. 

\item Are there any science drivers pushing for or against particular dithering patterns (either rotational dithers or translational dithers?)  If available, please mention specific simulated cadences, and specific metrics, that support your answer. 

\end{enumerate}

\section{Glossary of acronyms and abbreviations}

A table of acronyms and shorthands used in this paper and in general in LSST cadence-related work follows, to enhance the readability of this work and works within this Focus Issue, more definitions are available on the Rubin website.\footnote{\url{https://www.lsst.org/scientists/glossary-acronyms}}

\begin{longtable*}{ >{\raggedleft\arraybackslash}p{2cm}  |  p{4.1cm} | p{4.1cm} | p{3cm}   }  
\caption{Table of Acronyms and Shorthands}\\
\textbf{} & \textbf{Extended name} & \textbf{Description} & \textbf{URL and references} \\
\hline
\endfirsthead
\multicolumn{4}{c}%
{\tablename\ \thetable\ -- \textit{Continued from previous page}} \\
\textbf{} & \textbf{Extended name} & \textbf{Description} & \textbf{URL and references} \\
\hline
\endhead
\hline \multicolumn{4}{r}{\textit{Continued on next page}} \\
\endfoot
\hline
\endlastfoot

\hline  

\textbf{Rubin} & Vera C. Rubin Observatory &  Rubin is used as a shorthand to refer to the physical observatory, collection of software infrastructure including data reduction pipelines and software that enables cadence considerations, and the observatory employees & \url{https://www.lsst.org} \\

\textbf{LSST} & Legacy Survey or Space and Time &  The 10-year survey, or, more explicitly, collection of surveys including the WFD, DDF, and minisurveys, that will be performed over the first 10-years of life of Rubin Observatory. LSST was formerly the acronyms to refer to the Rubin Observatory Project by its earlier name of \emph{Large Synoptic Survey Telescope} & \url{https://www.lsst.org}\\

\textbf{SRD} & Survey Requirement Document & Core Rubin document defining the survey requirement that will enable the pursuit of the four core science goals of LSST (see \autoref{sec:intro}) & \url{https://ls.st/srd}\\

\textbf{WFD} & Wide Fast Deep & The main survey performed within LSST which is designed to enable, in conjunction with DDFs, the pursuit of the four core science goals of LSST (see \autoref{sec:intro})& \citet[][Section 3.1]{lsst}\\

\textbf{DDF} & Deep Drilling Fields & Selected single pointings in the southern hemisphere that will be observed at a denser observing cadence reaching higher magnitude limit in stacked images &  \url{https://www.lsst.org/scientists/survey-design/ddf}\\

\textbf{OpSim} & Operation Simulator & Rubin Software that enables the simulation of 10-year survey pointing databases, including observing strategy, whether, and telescope maintenance and operation constrains & \url{https://community.lsst.org/t/from-opsim-v4-to-fbs-1-2/3856}; \citet{2019AJ....157..151N, 2016SPIE.9910E..13D, 2014SPIE.9150E..15D}\\

\textbf{MAF} & Metric Analysis Framework & Rubin Software API that enables interacting with the databases containing simulated observing histories generated through \opsim &  \url{https://github.com/LSST-nonproject/sims_maf_contrib}; \citet{2014SPIE.9149E..0BJ}\\

\textbf{SAC} & Science Advisory Committee & Comprised of scientists familiar with but external to the Rubin Project, the SAC advises the Rubin Constructions and Operations Directors on both policy questions and technical topics of interest to the Project and the science community & \url{https://project.lsst.org/groups/sac/welcome}\\ 

\textbf{SCOC} & Survey Cadence Optimization Committee (SCOC) & Advisory body to the Rubin Observatory 
Operations Director formed in 2020,  responsible for optimizing the LSST cadence (see \autoref{ss:soc}) & \url{https://ls.st/scoc}\\ 

\textbf{SCs} & Rubin LSST Science Collaborations & Self-governed teams independent of Rubin Observatory but recognized by the SAC that come together to address specific scientific challenges and focus on specific science themes and their pursuit through LSST & \url{https://www.lsstcorporation.org/science-collaborations}\\ 

\textbf{AGN SC} & AGN Science Collaborations & SC pursuing Active Galactic Nuclei studies & \url{https://agn.science.lsst.org}\\ 

\textbf{DESC} & Dark Energy Science Collaboration & SC pursuing dark energy studies & \url{https://lsstdesc.org}\\ 

\textbf{Galaxies SC} & Galaxies Science Collaborations & SC addressing studies of galaxies and galaxy evolution & \url{https://sites.google.com/view/lsstgsc/home}\\ 

\textbf{ISSC} & Informatics and Statistics Science Collaboration & SC addressing methodological challenges specifit to the LSST data & \url{https://issc.science.lsst.org}\\ 

\textbf{SLSC} & Strong Lensing Science Collaborations & SC addressing strong-lensing studies & \url{https://sites.google.com/view/lsst-stronglensing}\\ 

\textbf{SMWLV} & Stars, Milky Way, Local Volume Science Collaboration & SC addressing Milky Way and local volume studies & \url{https://milkyway.science.lsst.org}\\ 

\textbf{SSSC} & Solar System Science Collaborations & SC addressing Solar System studies & \url{https://lsst-sssc.github.io}\\ 

\textbf{TVS SC} & Transient and Variable Stars Science Collaboration & SC addressing the study of the variable and transient sky, both Galactic and extragalactic & \url{https://lsst-tvssc.github.io}\\

\textbf{COSEP} & Community Observing Strategy Evaluation Paper & Living document collecting observing strategy consideration and analyses through MAF, primarily written by the Science Collaborations & \url{https://github.com/LSSTScienceCollaborations/ObservingStrategy}; \citet{COSEP}\\

\textbf{LSSTC} & LSST Corporation & A not-for-profit 501(c)3 corporation formed to initiate the LSST Project and advance the science of astronomy and physics. LSSTC represents a consortium of nearly 40 institutional members, as well as 34 international contributors representing 23 countries. LSSTC will partner with NSF/AURA and DOE/SLAC in Rubin Observatory LSST operations and enable the exploitation of the Rubin Observatory Legacy Survey of Space and Time’s (LSST) data by advocating for and supporting LSST science & \url{https://www.lsstcorporation.org}\\
\end{longtable*}

%\bibliography{refs}

\end{document}